\documentclass[11pt,twocolumn]{scrartcl}
\usepackage[a4paper, top = 2cm, left = 1.5cm,right=1.5cm, bottom = 2cm]{geometry}
\usepackage[ngerman, english]{babel}
\usepackage[utf8]{inputenc}
\usepackage{amsmath}
\usepackage{amsfonts}
\usepackage{amssymb}
\usepackage{multirow}
\usepackage{MnSymbol}
\usepackage{graphicx}
\usepackage{subfigure}
\usepackage{array}
\usepackage{hyperref}
\usepackage[headsepline]{scrlayer-scrpage}
\usepackage{nicefrac}
\usepackage{booktabs}
\usepackage{multicol}
\usepackage{xcolor}
\usepackage{dsfont}
\usepackage{soul}
\usepackage{chemformula}
\usepackage[autocite = superscript,backend=biber,style=numeric-ind,sorting=none,maxbibnames=2,citestyle=numeric-comp]{biblatex}

\begin{document}
\twocolumn[
\begin{@twocolumnfalse}
\thispagestyle{empty}
\begin{center}
\Large{\textbf{Interlayer Raman modes in twisted bilayer TMDCs}}
\end{center}
%\vspace{0.3cm}
\begin{center}
Eileen Schneider,$^{1,}$\footnote{eileen.schneider@fau.de} Kenji Watanabe,$^{2}$ Takashi Taniguchi,$^{3}$ and Janina Maultzsch$^{1}$
\par
\vspace{0.2cm}
\small{$^{1}$\textit{Department of Physics, Chair of Experimental Physics,\\ Friedrich-Alexander-Universität Erlangen-Nürnberg, 91058 Erlangen, Germany}}\par
\small{$^{2}$\textit{Research Center for Electronic and Optical Materials,\\ National Institute for Materials Science, 1-1 Namiki, Tsukuba 305-0044, Japan}}\par
\small{$^{3}$\textit{Research Center for Materials Nanoarchitectonics,\\ National Institute for Materials Science,  1-1 Namiki, Tsukuba 305-0044, Japan}}
\end{center}

\begin{abstract}
Twisted bilayer two-dimensional transition-metal dichalgocenides (TMDCs) exhibit a range of novel phenomena such as the formation of moiré excitons and strongly correlated phases. The coupling between the layers is crucial for the resulting physical properties and depends not just on the twist angle but also on the details of the fabrication process. Here, we present an approach for the analysis of this interlayer coupling \textit{via} Raman spectroscopy. By exciting the \textit{C}-exciton resonance of the TMDCs, optical (high-frequency) interlayer Raman modes are activated that are too weak in intensity at excitation energies far below the \textit{C} exciton. This is due to the wave function of the \textit{C} exciton, which expands - in contrast to \textit{A} and \textit{B} excitons - over both layers and, therefore, couples the layers electronically. We present optical interlayer Raman modes in twisted 2L TMDCs and show that they can be used as a measure of the interlayer coupling between the individual layers.
\end{abstract}
\end{@twocolumnfalse}
]
%\maketitle

\begin{figure}[htbp!]
\centering
\includegraphics[scale=0.34]{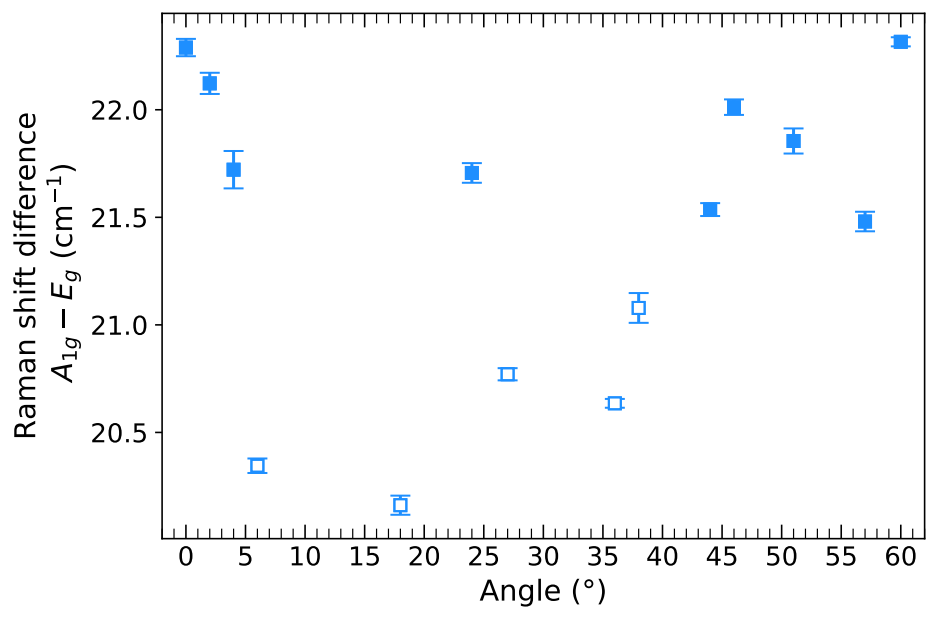}
\caption{\label{fig:MoS2ModeDiff} Frequency differences between the $A_{1g}$ and $E_{g}$ intralayer Raman modes of twisted 2L MoS$_2$ samples at different angles. The error bars are deduced from the Lorentz fits to the individual Raman peaks. Data points for samples that do not exhibit interlayer Raman modes are indicated with open squares as a guide to the eye. The Raman spectra were taken at a laser excitation wavelength of 532\,nm; spectra are displayed in Figure S2.}
\end{figure}

\cfoot{}
%\paragraph{Introduction.}
Transition metal dichalcogenides (TMDCs) have been studied intensely in recent years because of their physical properties being promisinig for applications like nanoelectronic and optoelectronic devices~\cite{Du2023}. Twisting two individual layers of TMDCs on top of each other gives a new degree of freedom to the system, namely the twist angle, which can lead to either commensurate or incommensurate structures. In commensurate structures, a new superlattice forms with a larger lattice constant than that of the single layer~\cite{Mak2022}. In both, commensurate and incommensurate twisted 2L systems, there may be areas with varying distance between the two constituent layers and local variations in the twist angle, due to imperfections during the fabrication process; in addition, the distance changes with twist angle as well~\cite{Mak2022}. This may influence the coupling between the two layers, which is an important precondition for a multitude of physical phenomena, such as correlated electronic phases~\cite{Wang2020,Regan2020,Balents2020}, superconductivity~\cite{An2020,Balents2020,Guo2024}, moiré and interlayer excitons~\cite{Jin2019,Miller2017,Zhao2024}.  All of these effects require sufficiently strong coupling without unwanted molecules between the layers or varying interlayer distance. However, a lack of coupling might only become apparent after numerous fabrication steps. Thus, a simple and fast method for validating the interlayer coupling in twisted 2L systems is highly desired in order to help understanding of new phenomena in twisted 2L systems.
\\\indent 
So far, the interlayer coupling in TMDC homo- or heterostructures has been investigated by Raman spectroscopy~\cite{Huang2014,Pan2022,Scheuschner2015,Plechinger2012}. In particular, the low-frequency interlayer shear mode has been used to examine the interlayer coupling, but only for twist angles near 0° and 60°~\cite{Parzefall2021,Puretzky2016}. In this work we show that optical interlayer Raman modes in twisted 2L TMDCs can be used as an indication for coupling between the two layers. We present Raman spectra of twisted 2L MoS$_2$ and MoSe$_2$ with twist angles ranging from 0° to 60°.  By varying the excitation energy, we show that the optical interlayer modes appear near the \textit{C}-exciton resonance, where the electronic wave functions are extended over both layers. In twisted 2L samples with insufficient coupling, the optical interlayer modes are not observed. This is confirmed by additional spectroscopic data, which point to single-layer like behavior in bilayers identified as uncoupled.
\\\indent 
Monolayers of MoSe$_2$ were mechanically exfoliated from a bulk crystal (HQ Graphene) onto PDMS and stacked onto each other by a dry-transfer method using polycarbonate as described in Ref.~\cite{Purdie2018}. A single flake is cut by a laser cutter and one half stamped onto the other. The twist angle can be controlled with a rotation stage and a 4-point angle measurement afterwards. Single layers of MoS$_2$ were grown \textit{via} chemical vapor deposition (CVD) on Si/SiO$_2$ (MicroChemicals) using molybdenum trioxide and sulfur powder (both from Sigma-Aldrich). The process was done at atmospheric pressure in argon atmosphere at a temperature between 740°C and 780°C. The single layers were delaminated from the growth substrate \textit{via} a water-assisted transfer method as shown in Ref.~\cite{Ma2017} and stamped on top of each other. The twist angle was measured between the edges of two triangular flakes. All twisted bilayers are encapsulated in hexagonal boron nitride (hBN) layers exfoliated from the bulk crystal~\cite{Taniguchi2007}. Raman measurements were conducted with a Horiba LabRam and a T64000 spectrometer at different laser wavelengths ranging from 405\,nm to 633\,nm, focused by a 100x objective. The laser power was kept below 0.5\,mW to avoid local heating. To compare Raman intensities at different laser energies, the intensities were calibrated using the silicon substrate and an additional calcium fluoride crystal (Crystal). The Raman shifts were calibrated by Neon lines.
\\\indent 
 \ohead{\thepage}
 \ihead{Schneider et al., Interlayer Raman modes in twisted bilayer TMDCs}
 A single-layer (1L) TMDC  (symmetry group D$_{3\text{h}}$) exhibits the characteristic $E'$ and $A_1'$ Raman modes with frequencies of about 383 and 407\,cm$^{-1}$  for MoS$_2$, respectively. The $E''$ mode at 286\,cm$^{-1}$ and $A_{2}''$ mode at 471\,cm$^{-1}$ are not observable due to selection rules~\cite{Scheuschner2015}. For each mode of the monolayer, there is one Davydov pair in two-layer (2L) and bulk TMDCs. One of these Davydov pairs is even under spatial inversion and, in this case, Raman active.  Consequently, the Raman-active modes in 2L TMDCs have $E_{g}$, $E_{g}$, $A_{1g}$ and $A_{1g}$ symmetry, besides the low-frequency rigid-layer modes~\cite{Scheuschner2015}. Since the $E''$ and $A_{2}''$ modes are not observable in 1L but their corresponding $E_{g}$ and $A_{1g}$ modes in 2L are, they are called optical interlayer modes. However, the intensity of these Raman modes depends strongly on the excitation energy. In resonance with the \textit{A} or \textit{B} exciton, a few-layer TMDC will rather behave as a superposition of independent layers because the \textit{A} and \textit{B} exciton wave functions are confined to a single layer. The \textit{C} exciton, on the other hand, expands over multiple layers, which leads to an electronic coupling of the layers~\cite{Gillen2017}. This leads to an increased intensity of these interlayer Raman modes and makes them observable in the Raman spectra. 
\\\indent 
Now the question arises whether optical interlayer Raman modes are visible in \textit{twisted} 2L TMDC systems as well, and whether they can be interpreted as an indication for coupling between the layers. In order to investigate the interlayer Raman modes in twisted TMDCs, we fabricated 14 different twisted  2L MoS$_2$ and five twisted 2L MoSe$_2$ samples as described above. They are each encapsulated in layers of hBN. A microscope image of a twisted 2L MoSe$_2$ sample is shown in Figure S1 in the supplementary information.
\\\indent 
 \autoref{fig:MoS2DiffAngles} shows Raman spectra of multiple twisted 2L MoS$_2$ samples excited just above the \textit{C}-exciton resonance at 405\,nm (3.06\,eV). The 0° sample is a naturally CVD-grown bilayer corresponding to 3R configuration; the 60° sample is an exfoliated MoS$_2$ bilayer corresponding to 2H configuration. The most prominent peaks are the characteristic $E_{g}$ and $A_{1g}$ modes of 2L MoS$_2$ at frequencies of about 385\,cm$^{-1}$ and 404\,cm$^{-1}$, respectively. In addition, folded phonon modes with frequencies depending on the size of the moiré Brillouin zone and, thus, the twist angle, are expected~\cite{Lin2018,Li2023}. Such a so-called $FA_{1g}$ mode at around 410\,cm$^{-1}$ is indeed observable in our samples for twist angles between 20° and 45° at an excitation wavelength of 532\,nm (see Figure S2 in the supplementary information).
 When excited at 405\,nm above the \textit{C}-exciton resonance, we observe in most samples two additional peaks at 286\,cm$^{-1}$ and 471\,cm$^{-1}$, which we assign to the $E_{g}$ and $A_{1g}$ optical interlayer Raman modes, respectively.
%\par %Results and Discussion
\begin{figure}[htbp!]
\centering
\includegraphics[scale=0.6]{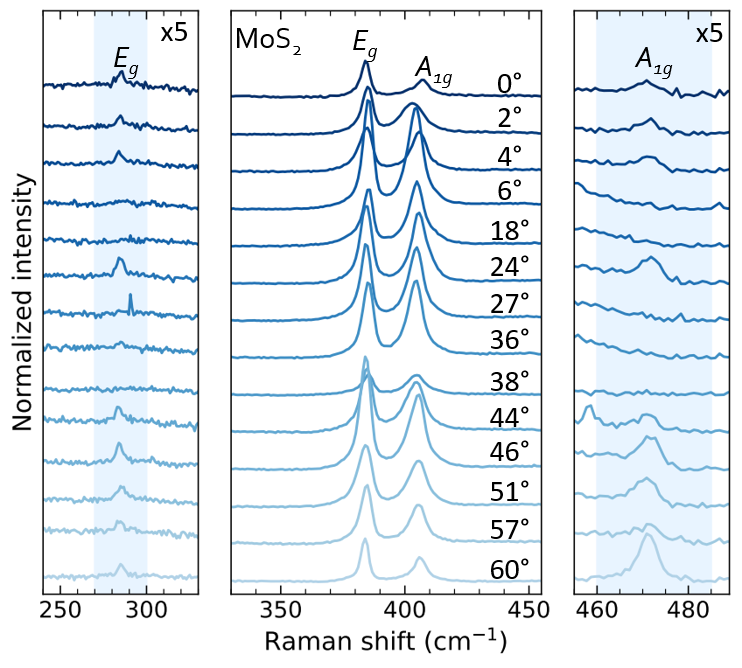}
\caption{\label{fig:MoS2DiffAngles} Raman spectra (laser wavelength 405\,nm) of twisted 2L MoS$_2$ with different twist angles given next to the spectra. The interlayer Raman modes at 286 and 471\,cm$^{-1}$ (left and right panels) are observable in some of the samples only. The intensities were calibrated using the Si substrate and CaF$_2$. The parts of the spectra in the left and right panels were multiplied by a factor of 5 for better visibility.}
\end{figure}
\\\indent 
 Since these phonon modes can only be activated in a bilayer system, we conclude that the presence of these modes in the Raman spectra indeed indicates coupling between the two layers mediated by the electronic wavefunctions. If the interlayer modes do not appear, the coupling must be weak, i.e., the twisted 2L structure rather corresponds to two individual layers. To further investigate this assumption, we measured the frequency difference between the intralayer $A_{1g}$ and $E_{g}$ Raman modes, which is typically used to determine the layer number~\cite{Lee2010}. \autoref{fig:MoS2ModeDiff} shows the ($A_{1g}-E_{g}$) Raman shift differences for all measured samples. For those samples that exhibit interlayer Raman modes, the data points are shown with closed squares, while the samples without interlayer Raman modes are displayed with open squares. For all samples without interlayer Raman modes, the frequency difference between the intralayer Raman modes is lower than for samples where interlayer Raman modes are observed. As the frequency difference is lowest for monolayers, this supports our assumption that the absence of optical interlayer Raman modes indeed points to a weak interlayer coupling, i.e., monolayer behavior.
\par
In order to further validate our conclusion that only those twisted 2L are electronically coupled, in which the optical interlayer Raman modes can be activated near the \textit{C}-exciton resonance, we performed photoluminescence (PL) measurements. For samples that do not exhibit interlayer Raman modes most PL spectra show an increase in the \textit{A} exciton peak intensity from 1L to 2L MoS$_2$, which supports the conclusion that these samples indeed behave like a stack of two separate layers. For samples with interlayer Raman modes, the intensity of the \textit{A} exciton peak in the monolayer is higher than that of the twisted bilayer, as is expected also for natural bilayers~\cite{Mak2010}. For exemplary PL spectra of two MoS$_2$ samples, see Figure S3 in the supplementary information. Our measurements suggest that interlayer Raman modes can be used to determine the coupling between the two constituent layers of twisted 2L TMDC systems. The coupling between two TMDC layers has also been investigated using rigid layer modes at low frequencies, namely the shear and breathing modes at around 20 and 40\,cm$^{-1}$, respectively~\cite{Parzefall2021,Puretzky2016,Huang2016}. These modes also appear only in few-layer systems and can already be observed for excitation energies below the \textit{C}-exciton resonance. However, the shear mode decreases in frequency for large twist angles between 20° and 40°, and is out of range for Raman spectroscopy; the breathing mode is very broad and therefore difficult to interpret confidently. Here, interlayer Raman modes offer a more distinct possibility for the investigation of interlayer coupling.
\\\indent 
Based on the results for twisted 2L MoS$_2$, we expect that the optical interlayer Raman modes indicate the coupling also in other 2L TMDCs, when excited near their \textit{C}-exciton resonance. In MoSe$_2$, the \textit{C} exciton is at about 2.4\,eV~\cite{Gillen2017}, thus, the interlayer Raman modes should be observed already at lower excitation energies than for MoS$_2$. 
%
%\begin{wrapfigure}{c}{1\textwidth}
\begin{figure}[htbp!]
\centering
\includegraphics[scale=0.8]{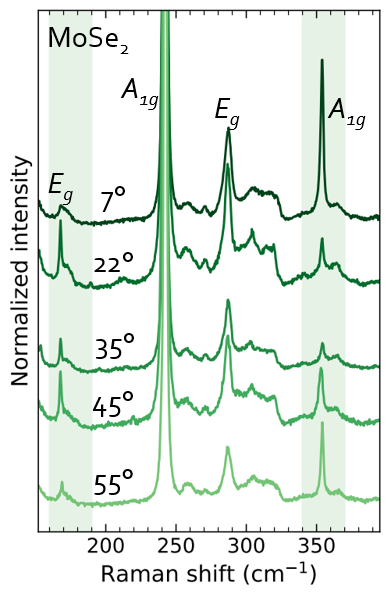}
\caption{\label{fig:MoSe2DiffAngles} Raman spectra of twisted 2L MoSe$_2$ (laser wavelength 514\,nm) with different twist angles given next to the spectra. The interlayer Raman modes at 169 and 353\,cm$^{-1}$ are observable in all samples. The intensities were calibrated using the Si substrate and CaF$_2$.}
%\end{wrapfigure}
\end{figure}
\par
\begin{figure}[htbp!]
\centering
\includegraphics[scale=0.36]{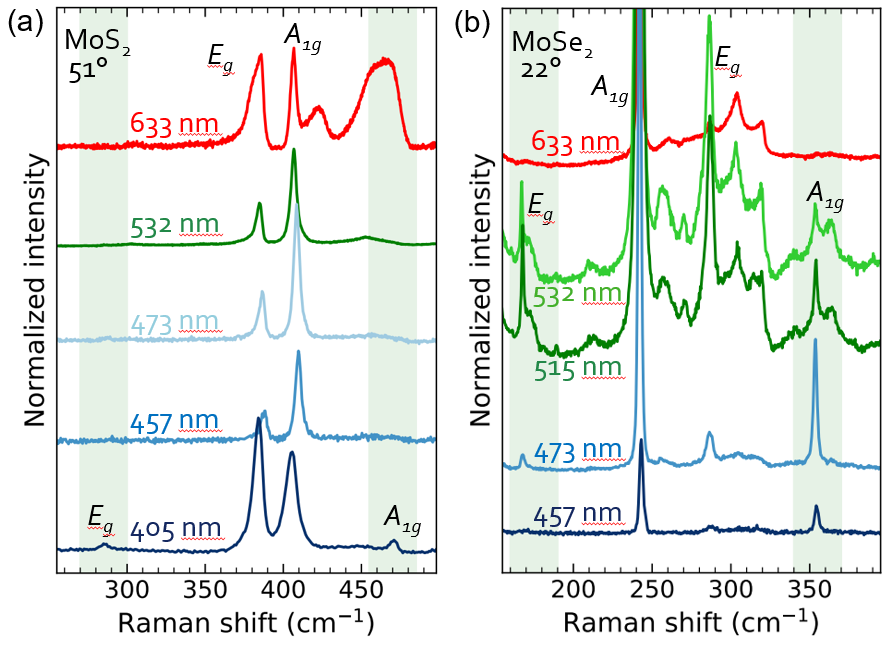}
\caption{\label{fig:MoS2DiffLaser} (a) Raman spectra of twisted 2L MoS$_2$ with a 51° twist angle and (b) twisted 2L MoSe$_2$ with a twist angle of 22° excited at different laser wavelengths given next to the spectra. In twisted 2L MoS$_2$, the interlayer Raman modes become visible only for excitation at 405\,nm, above the \textit{C} exciton in 2L MoS$_2$. In twisted 2L MoSe$_2$, the interlayer Raman modes become visible already at 532\,nm, which is near the \textit{C} exciton of 2L MoSe$_2$. The peak intensities were calibrated using the Si substrate and CaF$_2$.}
\end{figure}

In \autoref{fig:MoSe2DiffAngles} we show Raman spectra of multiple twisted 2L MoSe$_2$ samples with different twist angles excited at 514\,nm (2.41\,eV). Again, the two characteristic modes of MoSe$_2$ are prominent in the Raman spectra at 242\,cm$^{-1}$ and 285\,cm$^{-1}$ for the $A_{1g}$ and the $E_{g}$ mode, respectively~\cite{Nam2015}. Indeed, we observe the optical interlayer Raman modes at 169\,cm$^{-1}$ ($E_{g}$) and 353\,cm$^{-1}$ ($A_{1g}$) for all twist angles. Their frequencies are in accordance with literature for natural 2L MoSe$_2$~\cite{Nam2015,Soubelet2016}. In contrast to the twisted 2L MoS$_2$ samples, we observe interlayer Raman modes in all samples. This can be explained by the different fabrication methods: For the CVD-grown MoS$_2$ layers, a water-assisted transfer was used to delaminate the individual monolayers from their growth substrates and stamp them on top of each other. This might occasionally lead to water confined between the MoS$_2$ layers, which prevents the interlayer coupling. For the preparation of MoSe$_2$ twisted bilayers, no water was used in the fabrication process.
\\\indent 
In order to confirm the excitation-energy dependence of the optical interlayer modes also for twisted bilayers, we performed Raman measurements at different laser excitation energies. \autoref{fig:MoS2DiffLaser} shows Raman spectra of twisted 2L MoS$_2$ and MoSe$_2$ excited at several laser wavelengths. For MoS$_2$, the $E_g$ and $A_{1g}$ interlayer Raman modes only appear at excitation at 405\,nm (3.06\,eV), which is above the \textit{C}-exciton energy of about 2.8\,eV~\cite{Gillen2017}. The \textit{C}-exciton resonance of MoSe$_2$ lies at about 2.4\,eV~\cite{Gillen2017}. Indeed, in MoSe$_2$, the $E_{g}$ and $A_{1g}$ interlayer Raman modes are already observable at an excitation wavelength of 532\,nm (2.33\,eV) and become stronger at higher energies. 
\\ \indent 
In this work we have shown that the previously known optical interlayer Raman modes are not only present in few-layer TMDCs but also in twisted 2L TMDCs. The observed Raman peaks appear at excitation energies corresponding to the \textit{C}-exciton resonance in twisted 2L MoS$_2$ and twisted 2L MoSe$_2$ systems. Furthermore, these modes are only observed if the twisted layers are electronically coupled. This is supported by the $A_{1g} - E_{g}$ intralayer mode frequency differences and PL spectra of twisted 2L MoS$_2$ samples. These showed monolayer characteristics for all samples that did not exhibit interlayer Raman modes. Resonance Raman spectroscopy can, thus, be used as a fast method to investigate the interlayer coupling in twisted TMDC bilayers making use of the spatial extent of the electronic wave functions contributing to the \textit{C} exciton. 
\\
E.S. and J.M. acknowledge funding by the Dr. Isolde Dietrich-Stiftung and support by the Deutsche Forschungsgemeinschaft (DFG, German Research Foundation), project number 447264071 (INST 90/1183-1 FUGG). K.W. and T.T. acknowledge support from the JSPS KAKENHI (Grant Numbers 21H05233 and 23H02052) and World Premier International Research Center Initiative (WPI), MEXT, Japan.
\pagebreak
\printbibliography

\end{document}